\documentclass[preprint,nofootinbib,floats,superscriptaddress,tightenlines,amsfonts,amsmath,amssymb]{revtex4}
\pdfoutput=1

\usepackage{graphicx}
\usepackage{bm}
\usepackage{color}
\usepackage{hyperref}
\usepackage{multirow}
\usepackage{slashed}

\begin{document}
\baselineskip=17pt \parskip=5pt

\preprint{NCTS-PH/1815}\hspace*{\fill}

\title{Rare nonleptonic $\bm{\bar{B}_s^0}$ decays as probes of new physics\\
behind $\bm{b\to s\mu\bar\mu}$ anomalies}

\author{Gaber Faisel}
\email{gaberfaisel@sdu.edu.tr}
\affiliation{Department of Physics, Faculty of Arts and Sciences,
S\"uleyman Demirel University, D685, Isparta 32260, Turkey}

\author{Jusak Tandean}
\email{jtandean@yahoo.com}
\affiliation{Department of Physics, National Taiwan University, Taipei 10617, Taiwan}
\affiliation{Physics Division, National Center for Theoretical Sciences, Hsinchu 30013, 
Taiwan\bigskip}


\begin{abstract}

The anomalous results of recent measurements on various $b\to s\mu^+\mu^-$ processes could be 
initial evidence of physics beyond the standard model (SM).
Assuming this to be the case, we entertain the possibility that the underlying new physics also
affects the rare nonleptonic decays of the $\bar B_s^0$ meson.
We consider in particular new physics arising from the interactions of a heavy $Z'$ boson and 
investigate their influence on the decay modes $\bar B_s^0\to(\eta,\eta',\phi)\omega$, 
which receive sizable QCD- and electroweak-penguin contributions.
These decays are not yet observed, and their rates are estimated to be relatively small in the SM.
Taking into account the pertinent constraints, we find that the $Z'$ effects can greatly increase
the rates of $\bar B_s^0\to(\eta,\phi)\omega$, by as much as two orders of magnitude,
with respect to the SM expectations.
We have previously shown that $\bar B_s^0\to(\eta,\phi)\pi^0$, with similarly suppressed SM rates, 
could also undergo substantial $Z'$-induced enhancement.
These rare modes can therefore serve as complementary probes of the potential new physics which
may be responsible for the $b\to s\mu^+\mu^-$ anomalies.

\end{abstract}

\maketitle

\section{Introduction \label{sec:intro}}

The current data on a number of \,$b\to s\mu^+\mu^-$\, transitions have manifested several
tantalizing deviations from the expectations of the standard model (SM).
Specifically, the LHCb Collaboration\,\,\cite{Aaij:2015oid} found moderate tensions with the SM
in an angular analysis of the decay $B^0\to K^{*0}\mu^+\mu^-$, which were later corroborated in
the Belle experiment \cite{Wehle:2016yoi}.
Moreover, LHCb reported\,\,\cite{Aaij:2014ora} that the ratio $R_K$ of the branching
fractions of $B^+\to K^+\mu^+\mu^-$ and $B^+\to K^+e^+e^-$ decays and the corresponding ratio
$R_{K^*}$ for $B^0\to K^{*0}\mu^+\mu^-$ and $B^0\to K^{*0}e^+e^-$ decays are a couple of sigmas
below their SM predictions\,\,\cite{Hiller:2003js}.
Also, the existing measurements\,\,\cite{Aaij:2014pli,pdg} on the branching fractions
of $B\to K^{(*)}\mu^+\mu^-$ and $B_s\to\phi\mu^+\mu^-$ favor values below their SM estimates.

These anomalies may be harbingers of physics beyond the SM, although their statistical
significance is still insufficient for drawing a definite conclusion.
In fact, model-independent theoretical studies have pointed out that new physics (NP)
could explain them \cite{Capdevila:2017bsm,DAmico:2017mtc}.
This would suggest that they might be experimentally confirmed in the near future to have
originated from beyond the SM.
Thus, it seems timely to explore what might happen if the same underlying NP could have 
an appreciable influence on some other \,$b\to s$\, processes.

Previously we have entertained such a possibility in a scenario where a new, electrically neutral 
and uncolored, spin-one particle, the $Z'$ boson, is behind the \,$b\to s\mu^+\mu^-$\, 
anomalies \cite{Faisel:2017glo}.
In particular, we investigated the potential implications for the nonleptonic decays of
the $\bar B_s^0$ meson which are purely isospin-violating, namely
\,$\bar B_s^0\to(\eta,\eta',\phi)(\pi^0,\rho^0)$,\, most of which are not yet observed \cite{pdg}.
In the SM limit, they are not affected by QCD-penguin operators, which conserve
isospin, while the effects of tree operators are suppressed by a factor
\,$|V_{us}V_{ub}|/|V_{ts}V_{tb}|\sim0.02$\, involving Cabibbo-Kobayashi-Maskawa (CKM) matrix 
elements, and thus the amplitudes for these decays tend to be dominated by electroweak-penguin
contributions \cite{Fleischer:1994rs}.
Accordingly, their rates in the SM are comparatively small \cite{Fleischer:1994rs,Tseng:1998wm,
Cheng:2009mu,Ali:2007ff,Yan:2018fif,Yan:2017nlj,Wang:2017rmh,Wang:2008rk,Williamson:2006hb,
Faisel:2011kq}, which motivated earlier works suggesting that one or more of these decay modes
could be sensitive to NP signals \cite{Faisel:2011kq,Hofer:2010ee,Hua:2010wf,Chang:2013hba,
Faisel:2014dna,Bobeth:2014rra}.
Incorporating the relevant constraints, we demonstrated in Ref.\,\cite{Faisel:2017glo} that
the $Z'$ influence could cause the rates of two of the modes, \,$\bar B_s^0\to(\eta,\phi)\pi^0$,\,
to rise by up to an order of magnitude above their SM expectations.
It follows that these modes could offer valuable complementary information about the NP
which may be responsible for the \,$b\to s\mu^+\mu^-$\, anomalies.

Extending our preceding analysis, the present paper covers \,$\bar B_s\to(\eta,\eta',\phi)\omega$,\, 
which are also not yet observed \cite{pdg}.
Here, as in the \,$\bar B_s^0\to(\eta,\eta',\phi)(\pi^0,\rho^0)$\, case, the tree operators suffer 
from the CKM suppression, again allowing the penguin operators to become important.
However, unlike the latter modes, \,$\bar B_s\to(\eta,\eta',\phi)\omega$\, preserve isospin and
therefore receive both electroweak- and QCD-penguin contributions.
In the SM, the rates of these decays turn out to be relatively small as
well \cite{Cheng:2009mu,Ali:2007ff,Yan:2018fif,Yan:2017nlj,Wang:2017rmh,Wang:2008rk}, and so they 
could be expected to serve as additional probes of the potential NP behind the anomalies.
We will show that this can indeed be realized in the aforementioned $Z'$ model,
especially for the two modes \,$\bar B_s\to(\eta,\phi)\omega$.\,

The remainder of the paper is organized as follows.
In Sec.\,\,\ref{ffz'} we describe the $Z'$ interactions which impact the various processes of concern.
In Sec.\,\,\ref{Bs2MM'} we address how the $Z'$-induced effects on the considered rare nonleptonic 
decays could raise some of their rates significantly.
We will impose appropriate restraints on the $Z'$ couplings, including from other \,$b\to s$\, 
transitions, such as \,$\bar B_s\to\phi\rho^0$\, and \,$B\to\pi  K$\, decays.\footnote{The possibility 
of NP in $b\to s\mu^+\mu^-$ producing detectable implications for the \,$B\to\pi  K$\, decays has 
previously been brought up in \cite{Beaudry:2017gtw}.\medskip}
Our numerical work will also involve \,$\bar B_s\to(\eta,\phi)\pi^0$,\, which we investigated before, 
to see if there might be any correlation between their rate enlargement and that of 
\,$\bar B_s\to(\eta,\phi)\omega$.\,
In Sec.\,\,\ref{conclusions} we give our conclusions.
An Appendix contains extra formulas.

\section{\boldmath$Z'$ interactions\label{ffz'}}

In our $Z'$ scenario of interest, the mass eigenstates of the $u$, $d$, $s$, and $b$ quarks have
nonstandard interactions described by \cite{Faisel:2017glo}
\begin{align} \label{LsbZ'}
{\mathcal L}_{Z'}^{} \,\supset &\,~
\mbox{$-\big[$} \overline{s}\,\gamma^\kappa\big(\Delta_L^{sb}P_L^{}+\Delta_R^{sb}P_R^{}\big)b\,Z_\kappa'
\;+\; {\rm H.c.} \big] \,-\, \Delta_V^{\mu\mu}\,\overline{\mu}\,\gamma^\kappa\mu\,Z_\kappa'
\nonumber \\ & -\,
\big[ \overline{u}\,\gamma^\kappa\big(\Delta_L^{uu}P_L^{}+\Delta_R^{uu}P_R^{}\big)u +
\overline{d}\,\gamma^\kappa\big(\Delta_L^{dd}P_L^{}+\Delta_R^{dd}P_R^{}\big)d \big]Z_\kappa' \,,
\end{align}
where the constants $\Delta_{L,R}^{sb}$ are generally complex, while $\Delta_V^{\mu\mu}$ and 
$\Delta_{L,R}^{uu,dd}$ are real due to the Hermiticity of ${\mathcal L}_{Z'}$, and
\,$P_{L,R}=(1\mp\gamma_5^{})/2$.\,
As in Ref.\,\,\cite{Faisel:2017glo}, we suppose that any other possible couplings of the $Z'$ to SM
fermions are negligible and that it does not mix with SM gauge bosons but is not necessarily
a gauge boson.\footnote{In the literature pertaining to the anomalies, different $Z'$ models have 
been explored, some of which can be found in \cite{b2sll-list}.}
Moreover, for simplicity we concentrate on the special case in which
\begin{equation} \label{Dsb}
\Delta_L^{sb} \,=\, \lambda_t^{}\, \rho_L^{} \,, ~~~~ ~~~
\Delta_R^{sb} \,=\, \lambda_t^{}\, \rho_R^{} \,, ~~~~ ~~~ ~~
\lambda_q^{} \,=\, V_{qs}^*V_{qb}^{} \,,
\end{equation}
where $\rho_{L,R}^{}$ are real numbers.

For the $Z'$ being heavy, the couplings to $b\bar s$ and $\mu\bar\mu$ in Eq.\,(\ref{LsbZ'})
contribute to the effective Lagrangian
\begin{equation} \label{Lb2smm}
{\mathcal L}_{\rm eff}^{} \,\supset\, \frac{\alpha_{\rm e}^{}\lambda_t^{}G_{\rm F}^{}}
{\sqrt2\,\pi} \big( C_{9\mu}^{}\, \overline{s}_{\,}\gamma^\kappa P_L^{} b + C_{9'\mu}^{} \,
\overline{s}_{\,}\gamma^\kappa P_R^{} b \big) \overline{\mu}_{\,}\gamma_\kappa^{}\mu
\;+\; {\rm H.c.} \,,
\end{equation}
where $\alpha_{\rm e}^{}$ and $G_{\rm F}$ are the usual fine-structure and Fermi constants,
and \,$C_{9\mu}^{}=C_{9\ell}^{\textsc{sm}}+C_{9\mu}^{\textsc{np}}$ and
\,$C_{9'\mu}=C_{9'\mu}^{\textsc{np}}$ are the Wilson coefficients,
with $C_{9\ell}^{\textsc{sm}}$ being the flavor-universal SM part ($\ell=e,\mu,\tau$)
and \cite{Faisel:2017glo}
\begin{equation} \label{c9c9'}
C_{9\mu}^{\textsc{np}} \,=\, \frac{-\sqrt2\,\pi\,\rho_L^{}\Delta_V^{\mu\mu}}
{\alpha_{\rm e}^{}G_{\rm F}^{}\,m_{Z'}^2} \,, ~~~~~~~
C_{9'\mu}^{\textsc{np}} \,=\, \frac{-\sqrt2\,\pi\,\rho_R^{}\Delta_V^{\mu\mu}}
{\alpha_{\rm e}^{}G_{\rm F}^{}\,m_{Z'}^2} \,.
\end{equation}
According to model-independent analyses \cite{Capdevila:2017bsm},
one of the best fits to the anomalous \,$b\to s\mu^+\mu^-$\, data
corresponds to \,$C_{9\mu}^{\textsc{np}}\sim-1.1$\, and
\,$C_{9'\mu}^{\textsc{np}}\sim0.4$,\, with no NP in \,$b\to
se^+e^-$.\,

The $bsZ'$ couplings in ${\mathcal L}_{Z'}$ above also affect $B_s$-$\bar B_s$ mixing
at tree level and hence need to satisfy the restrictions inferred from its data.
As elaborated in Ref.\,\,\cite{Faisel:2017glo}, the requirements from \,$b\to s\mu^+\mu^-$\,
processes and $B_s$-$\bar B_s$ mixing together imply that the left-handed $bsZ'$ coupling must be
roughly ten times stronger than the right-handed one, and so \,$\rho_L^{}\sim10\rho_R^{}$.\,
This will be taken into account later on.

Additionally, ${\mathcal L}_{Z'}$ can yield modifications to nonleptonic
transitions, such as \,$\bar B_s\to(\eta,\eta',\phi)\,\omega $.\,
In the SM, their amplitudes proceed from \,$b\to s$\, four-quark operators $O_{1,2}^u$,
$O_{3,4,5,6}$, and $O_{7,8,9,10}$ derived from charmless tree, QCD-penguin, and
electroweak-penguin diagrams, respectively.\footnote{The expressions for
$O_j$, \,$j=1,2,\cdots,10$,\, are available from, {\it e.g.}, \cite{Williamson:2006hb}.}
In models beyond the SM, new ingredients may alter the Wilson coefficients $C_j$ of $O_j$
and/or generate extra operators $\tilde O_j$ which are the chirality-flipped
counterparts of $O_j$.
In our $Z'$ case, at the $W$-mass ($m_W^{}$) scale only $C_{3,5,7,9}$ and $\tilde C_{3,5,7,9}$
get $Z'$ contributions given by \cite{Barger:2009qs,Hua:2010wf,Hofer:2010ee,Chang:2013hba}
\begin{align}
{\cal L}_{\rm eff}^{} \,\supset&~ \sqrt8\,\lambda_t^{~}G_{\rm F}^{}
\raisebox{3pt}{\footnotesize$\displaystyle\sum_{q=u,d}$}
\Big\{ \overline{s}\,\gamma^\kappa P_L^{}b
\Big[ \Big( C_3^{}+\tfrac{3}{2}C_9^{}e_q^{} \Big) \overline{q}\,\gamma_\kappa^{}P_L^{} q
+ \Big( C_5^{}+\tfrac{3}{2}C_7^{}e_q^{} \Big) \overline{q}\,\gamma_\kappa^{}P_R^{} q \Big]
\nonumber \\ & \hspace{1in} +\,
\overline{s}\,\gamma^\kappa P_R^{}b \Big[ \Big( \tilde C_3^{}
+ \tfrac{3}{2}\tilde C_9^{}e_q^{} \Big) \overline{q}\,\gamma_\kappa^{}P_R^{} q
+ \Big( \tilde C_5^{}+\tfrac{3}{2}\tilde C_7^{}e_q^{} \Big)
\overline{q}\,\gamma_\kappa^{}P_L^{} q \Big] \Big\} \,,
\end{align}
where \,$C_i^{}=C_i^{\textsc{sm}} + C_i^{Z'}$\, and \,$\tilde C_i^{}=\tilde C_i^{Z'}$\, for
\,$i=3,5,7,9$\, are the Wilson coefficients with \cite{Faisel:2017glo}
\begin{align} \label{cz'}
C_{3,5}^{Z'} & \,=\, \frac{\rho_L^{}\bigl(-\delta_{L,R}^{}-3\Delta_{L,R}^{dd}\big)}
{6\sqrt2\,G_{\rm F}^{}\,m_{Z'}^2} \,, &
C_{7,9}^{Z'} & \,=\, \frac{-\rho_{L\,}^{}\delta_{R,L}^{}}{3\sqrt2\,G_{\rm F}^{}\,m_{Z'}^2} \,, &
\nonumber \\
\tilde C_{3,5}^{Z'} & \,=\, \frac{\rho_R^{}\bigl(-\delta_{R,L}^{}-3\Delta_{R,L}^{dd}\big)}
{6\sqrt2\,G_{\rm F}^{}\,m_{Z'}^2} \,, &
\tilde C_{7,9}^{Z'} & \,=\,
\frac{-\rho_{R\,}^{}\delta_{L,R}^{}}{3\sqrt2\,G_{\rm F}^{}\,m_{Z'}^2} \,, &
\\ \label{dLdR} \vphantom{|^{\int_\int^\int}}
\delta_L^{} & \,=\, \Delta_L^{uu} - \Delta_L^{dd} \,, &
\delta_R^{} & \,=\, \Delta_R^{uu} - \Delta_R^{dd} \,, &
\end{align}
we have assumed that renormalization group evolution (RGE) between the $m_{Z'}^{}$
and $m_W^{}$ scales can be neglected, and \,$e_u^{}=-2e_d^{}=2/3$.\,
At the $b$-quark mass ($m_b^{}$) scale, all the penguin coefficients acquire $Z'$ terms via
RGE, which we treat in the next section.

\section{\boldmath$Z'$ effects on rare nonleptonic $\bar B_s$ decays\label{Bs2MM'}}

To estimate the $Z'$ impact on \,$\bar B_s\to(\eta,\eta',\phi)\,\omega $,\,
following Ref.\,\cite{Faisel:2017glo} we employ the soft-collinear effective theory
(SCET)\,\,\cite{Wang:2017rmh,Wang:2008rk,Williamson:2006hb,Bauer:2000yr}.
For any one of them, we can write the SCET amplitude at leading order in the strong coupling 
$\alpha_s(m_b)$ as \cite{Wang:2008rk}
\begin{align} \label{AB2MM'}
{\cal A}_{\bar B_s\to M_1 M_2}^{} \,=&~\, \frac{f_{M_1}^{}G_{\rm F}^{}m_{B_s}^2}{\sqrt2} \bigg[
\int_0^1 \! d\nu \Big( \zeta_J^{BM_2\,} T_{1J}^{}(\nu) + \zeta_{Jg}^{BM_2\,} T_{1Jg}^{}(\nu)
\Big) \phi_{M_1}^{}(\nu)
+ \zeta^{BM_2\,} T_1^{} + \zeta_g^{BM_2\,} T_{1g}^{} \bigg]
\nonumber \\ & +\, (1\leftrightarrow 2) \;,
\end{align}
where $f_M^{}$ denotes the decay constant of meson $M$, the $\zeta$s are nonperturbative
hadronic parameters extractable from experiment, the $T$s represent hard kernels containing
the Wilson coefficients $C_j$ and $\tilde C_j$ at the $m_b^{}$ scale, and $\phi_{M}^{}(\nu)$
is the light-cone distribution amplitude of $M$ normalized as \,$\int_0^1d\nu\,\phi_M(\nu)=1$.\,
We collect the hard kernels, from Refs.\,\,\cite{Wang:2017rmh,Wang:2008rk,Williamson:2006hb}, 
in Table\,\,\ref{kernels}, where the flavor states \,$\eta_q\sim\big(u\bar u+d\bar d\big)/\sqrt2$\, 
and \,$\eta_s^{}\sim s\bar s$\, are linked to the physical meson states $\eta$ and $\eta'$ by
\,$\eta=\eta_q\cos\theta-\eta_s^{}\sin\theta$\, and
\,$\eta'=\eta_q\sin\theta+\eta_s^{}\cos\theta$\, with mixing angle \,$\theta=39.3^\circ$
\,\cite{Wang:2008rk,Williamson:2006hb,Feldmann:1998vh}.
We note that the so-called charming-penguin contribution is absent from
${\cal A}_{\bar B_s\to M_1 M_2}$, which is one of the reasons why these decays have low
rates \cite{Wang:2017rmh,Wang:2008rk}.

\begin{table}[b] \smallskip
\begin{tabular}{|c||c|c|c|c|} \hline
\,Decay mode\, & $~T_1~$ & $T_2$ & $~T_{1g}~$ & $T_{2g}\vphantom{\int_|^|}$ \\
\hline\hline
$\bar{B}_s\to\eta_s^{}\omega$ & 0 & ~$\frac{1}{\sqrt2}(c_2+c_3+2c_5+2c_6)$~ & 0 &
~$\frac{1}{\sqrt2}(c_2+c_3+2c_5+2c_6)\vphantom{\int_{\int_|}^{\int}}$~  \\
$\bar{B}_s\to\eta_q\omega$ & 0 & 0 & 0 & $c_2+c_3+2c_5+2c_6\vphantom{\int_{\int_|}^{\int}}$  \\
$\bar{B}_s\to\phi\omega$ & 0 &
$\frac{1}{\sqrt2}(c_2+c_3+2c_5+2c_6)\vphantom{\int_{\int_|}^{\int}}$ & 0 & 0 \\ \hline
\end{tabular}
\caption{Hard kernels $T_{1,2,1g,2g}$ for \,$\bar B_s\to(\eta,\eta',\phi)\omega$\, decays.
The hard kernels $T_{rJ,rJg}(\nu)$ for \,$r=1,2$\, are obtainable from $T_{r,rg}^{}$, respectively,
with the replacement \,$c_k^{}\to b_k^{}$,\, where $b_k^{}$ depends on $\nu$.} \label{kernels}
\end{table}

In the presence of NP which also gives rise to $\tilde O_j$, the quantities $c_k$ and $b_k$ in
Table\,\,\ref{kernels} depend not only on $C_j$ and $\tilde C_j$ but also on the final mesons
$M_{1,2}$ besides the CKM factors $\lambda_{u,t}$.
The dependence on $M_{1,2}$ is due to the fact that, in view of the nonzero kernels in this
table, for each 4-quark operator the contraction of the $\bar B_s$\,$\to$\,$M_1$ and
vacuum\,$\to$\,$M_2$ matrix elements in the amplitude can lead to an overall negative or positive
sign for the contribution of the operator, the sign being fixed by the chirality combination
of the operator and by whether $M_{1,2}$ are pseudoscalars ($PP'$), vectors ($VV'$), or $PV$.
Thus, for \,$\bar B_s\to PP'$\, and \,$\bar B_s\to\phi\omega$\,  we have
\begin{align} \label{cb}
c_2^{} & \,=\, \lambda_u^{} \Bigg( C_2^- + \frac{C_1^-}{N_{\rm c}} \Bigg)
- \frac{3\lambda_t}{2} \Bigg( C_9^- + \frac{C_{10}^-}{N_{\rm c}}\Bigg) , ~~~ ~~~~
c_3^{} \,=\, -\frac{3\lambda_t}{2} \Bigg( C_7^- + \frac{C_8^-}{N_{\rm c}} \Bigg) ,
\nonumber \\
c_{5,6}^{} & \,=\, -\lambda_t \Bigg( C_{3,5}^- + \frac{C_{4,6}^-}{N_{\rm c}}
- \frac{C_{9,7}^-}{2} - \frac{C_{10,8}^-}{2 N_{\rm c}} \Bigg) ,
\nonumber \\
b_2^{} & \,=\, \lambda_u^{} \Bigg[ C_2^- + \bigg(1-\frac{m_b^{}}{\omega_3}\bigg)
\frac{C_1^-}{N_{\rm c}} \Bigg] - \frac{3\lambda_t}{2} \Bigg[
C_9^- + \bigg(1-\frac{m_b^{}}{\omega_3}\bigg) \frac{C_{10}^-}{N_{\rm c}} \Bigg] ,
\nonumber \\
b_3^{} & \,=\, -\frac{3 \lambda_t}{2} \Bigg[ C_7^- + \bigg(1-\frac{m_b^{}}{\omega_2}\bigg)
\frac{C_8^-}{N_{\rm c}} \Bigg] ,
\nonumber \\
b_{5,6}^{} & \,=\, -\lambda_t \Bigg[ C_{3,5}^- + \bigg(1-\frac{m_b^{}}{\omega_3}\bigg)
\frac{C_{4,6}^-}{N_{\rm c}} - \frac{C_{9,7}^-}{2}
- \bigg(1-\frac{m_b^{}}{\omega_3}\bigg) \frac{C_{10,8}^-}{2 N_{\rm c}} \Bigg] ,
\end{align}
where \,$N_{\rm c}=3$\, is the color number, \,$C_j^-=C_j^{}-\tilde C_j^{}$,\,  and
$b_{2,3,5,6}$, which enter $T_{2J,2Jg}(\nu)$, are also functions of $\nu$ because
\,$\omega_2^{}=\nu m_{B_s}$\, and \,$\omega_3^{}=(\nu-1)m_{B_s}$\, \cite{Wang:2008rk}.
However, for \,$\bar{B}_s\to(\eta_q,\eta_s^{})\omega$\, we need to make the sign change
\,$C_j^-\to C_j^+=C_j^{}+\tilde C_j^{}$\, in $c_{2,3,5,6}$ and $b_{2,3,5,6}$.

The formulas in Eq.\,\eqref{cb} generalize the SM ones from Refs.\,\,\cite{Wang:2017rmh,Wang:2008rk,
Williamson:2006hb}, which also provide the $C_j^{\textsc{sm}}$ values at the $m_b$ scale, \,$C_{1,2}^{\textsc{sm}}=(1.11,-0.253)$\, and \,$C_{7,8,9,10}^{\textsc{sm}}=(0.09,0.24,-10.3,2.2)\times10^{-3}$,\, calculated at leading-logarithm order in the naive dimensional regularization scheme 
\cite{Buchalla:1995vs} with the prescription of Ref.\,\,\cite{Beneke:2001ev}.
We will incorporate these numbers into $c_k$ and $b_k$.
The $Z'$-generated coefficients in Eq.\,(\ref{cz'}) contribute to Eq.\,\eqref{cb} via \,$C_i^{}=C_i^{\textsc{sm}}+\delta C_i^{}$\, and \,$\tilde C_i^{}=\delta\tilde C_i^{}$\, for \,$i=3,4,...10$,\, 
where $\delta C_i^{}$ are linear combinations of $C_{3,5,7,9}^{Z'}$ due to RGE from the $m_W^{}$ scale 
to the $m_b^{}$ scale and $\delta\tilde C_i^{}$ are analogously related to $\tilde C_{3,5,7,9}^{Z'}$.

To evaluate ${\cal A}_{\bar B_s\to M_1 M_2}$, in light of Table\,\,\ref{kernels}, we employ the decay 
constant \,$f_\omega^{}=192$\,MeV\, \cite{Wang:2017rmh} and treat the integral in Eq.\,(\ref{AB2MM'}) 
with the aid of
\,$\int_0^1d\nu\,\phi_M^{}(\nu)/\nu=\int_0^1d\nu\,\phi_M^{}(\nu)/(1-\nu)\equiv
\langle\chi^{-1}\rangle_M^{}$\,
for \,$M=\eta_{q,s}^{},\phi$,\, in which cases \,$\langle\chi^{-1}\rangle_{\eta_{q,s}}=3.3$\, and
\,$\langle\chi^{-1}\rangle_\phi^{}=3.54$\, \cite{Wang:2008rk,Williamson:2006hb}.
Furthermore, for the $\zeta$'s in ${\cal A}_{\bar B_s\to(\eta_q,\eta_s)\omega}$, we adopt the two 
solutions from the fit to data performed in Ref.\,\,\cite{Wang:2008rk}:
\begin{align} \label{zetas}
\big(\zeta^P,\zeta_J^P,\zeta_g^{},\zeta_{Jg}^{}\big)_1 & \,=\, (0.137,0.069,-0.049,-0.027) \,,
\nonumber \\
\big(\zeta^P,\zeta_J^P,\zeta_g^{},\zeta_{Jg}^{}\big)_2 & \,=\, (0.141,0.056,-0.100,0.051) \,.
\end{align}
It is worth remarking here that, since they were the outcome of fitting to \,$B\to PP',PV$ data 
with SM Wilson coefficients \cite{Wang:2008rk}, using these solutions in an investigation of NP is 
justifiable provided that the impact of the NP on the channels which dominate the fit is small 
compared the SM contribution.
This requisite can be met by our $Z'$ scenario, as we set the mass of the $Z'$ to be $\cal O$(1\,TeV) 
and ensure that its couplings comply with the various constraints described in this study.
From Eq.\,(\ref{zetas}), we then have \,$\zeta_{(J)}^{B\eta_{q,s}}=\zeta_{(J)}^P$ and 
\,$\zeta_{(J)g}^{B\eta_{q,s}}=\zeta_{(J)g}^{}$,\, assuming flavor-SU(3) symmetry\,\,\cite{Wang:2008rk}.
Other input parameters are CKM matrix elements from Ref.\,\cite{utfit} as well as the meson masses 
\,$m_\eta=547.862$, \,$m_{\eta'}=957.78$, \,$m_\omega^{}=782.65$,\, \,$m_\phi^{}=1019.461$,\, and 
\,$m_{B_s}=5366.89$,\, all in units of MeV, and the $B_s$ mean lifetime 
$\tau_{B_s}=1.509\times10^{-12}$\,s,\, which are their central values from Ref.\,\,\cite{pdg}.
For the third $(\phi\omega)$ mode, we choose the CKM and SCET parameters supplied recently in 
Ref.\,\,\cite{Wang:2017rmh}.

\begin{table}[b] \medskip
\begin{tabular}{|c||c|c|c|c|} \hline
\multirow{2}{*}{$\begin{array}{c}\rm Decay \vspace{-3pt} \\ \rm mode \end{array}$} &
\multirow{2}{*}{QCDF} & \multirow{2}{*}{PQCD} & \multicolumn{2}{c|}{SCET} \\
\cline{4-5} & & & \small Solution 1 & \small Solution 2 \\ \hline\hline
$\bar B_s\to\eta\omega$ & $0.03_{-0.02-0.01}^{+0.12+0.06}\vphantom{|_{|_o}^{\int^o}}$ &
~$0.11_{-0.03}^{+0.04}$~ & $0.04_{-0.02}^{+0.04}$ & ~$0.007_{-0.002}^{+0.011}$~
\\
~$\bar B_s\to\eta'\omega$~  & $0.15_{-0.08-0.06}^{+0.27+0.15}\vphantom{|_{|_o}^{\int^o}}$ &
$0.35_{-0.04}^{+0.06}$ & ~$0.001_{-0.000}^{+0.095}$~ & $0.20_{-0.17}^{+0.34}$
\\ \hline\hline
$\bar B_s\to\phi\omega$  & ~$0.18_{-0.12-0.04}^{+0.44+0.47}$~ & \,$0.22_{-0.10}^{+0.15}$\, &
\multicolumn{2}{c|}{$0.04\pm0.01\vphantom{|_{|_o}^{\int^o}}$}
\\ \hline
\end{tabular}
\caption{Branching fractions, in units of $10^{-6}$, of \,$\bar B_s\to(\eta,\eta',\phi)\,\omega$\,
decays in the SM.
For the first two modes, the last two columns correspond to the two solutions of SCET parameters
in Eq.\,(\ref{zetas}).
The second and third columns exhibit numbers computed with QCDF \cite{Cheng:2009mu} and PQCD \cite{Yan:2017nlj,Yan:2018fif}.} \label{smbf} \vspace{-1ex}
\end{table}

Before dealing with the $Z'$ influence on \,$\bar B_s\to(\eta,\eta',\phi)\omega$\, numerically, with 
the above SCET prescription we arrive at their SM branching fractions, listed in Table\,\,\ref{smbf}.
For \,$\bar B_s\to(\eta,\eta')\omega$,\,  the entries in the last two columns correspond to the two 
solutions of SCET parameters in Eq.\,(\ref{zetas}).
The central values of the SCET predictions agree with those in Refs.\,\,\cite{Wang:2017rmh,Wang:2008rk}, 
from which we have added the errors shown.
For comparison, in the second and third columns we quote results from the QCD factorization (QCDF) 
\cite{Cheng:2009mu} and perturbative QCD (PQCD) \cite{Yan:2017nlj,Yan:2018fif} approaches.
Evidently, the SCET numbers can be compatible with the QCDF and PQCD ones within the sizable errors.
One concludes that for NP to be unambiguously noticeable in the rates it would have to amplify them 
relative to their SM ranges by significantly more than a factor of two.

In the presence of the $Z'$ contributions, we find the changes of the Wilson coefficients 
$C_j^{}$ at the $m_b$ scale to be
\begin{align}
\delta C_{1,2}^{} & \,=\, 0 \,, ~~~ \delta C_3^{} \,\simeq\, 1.13\, C_3^{Z'} \,, ~~~
\delta C_4^{} \,\simeq\, -0.29\, C_3^{Z'} \,, ~~~ \delta C_5^{} \,\simeq\, 0.93\, C_5^{Z'} \,, ~~~
\delta C_6^{} \,\simeq\, 0.29\, C_5^{Z'} \,,
\nonumber \\
\delta C_7^{} &  \,\simeq\, 0.93\, C_7^{Z'} \,, ~~~
\delta C_8^{}    \,\simeq\, 0.31\, C_7^{Z'} \,, ~~~
\delta C_9^{}    \,\simeq\, 1.11\, C_9^{Z'} \,, ~~~
\delta C_{10}^{} \,\simeq\, -0.25\, C_9^{Z'} \,,
\end{align}
where in each coefficient we have kept only the $Z'$ term with the biggest numerical factor,
upon applying the RGE at leading-logarithm order \cite{Buchalla:1995vs} with
\,$\alpha_e^{}=1/128$,\, $\alpha_{\rm s}^{}(m_Z^{})=0.119$,\, $m_b^{}=4.8$\,GeV,\, and
$m_t^{}=174.3$ GeV \,\cite{Williamson:2006hb}.
Furthermore, at the $m_b$ scale \,$\tilde C_j=\delta\tilde C_j$\, are related to 
$\tilde C_{3,5,7,9}^{Z'}$ in an analogous manner.
Combining the SM and $Z'$ portions, for \,$m_{Z'}^{}=1$ TeV and the central values of the input
parameters we derive the amplitudes for \,$\bar B_s\to(\eta,\eta',\phi)\,\omega$\, to be,
in units of $10^{-9}$ GeV,
\begin{align} \label{ampso}
{\cal A}_{\bar B_s\to\eta\omega}^{(1)} & \,\simeq \begin{array}[t]{l} -0.63 + 2.05 i
+ \big[ (6.06-0.12 i) \delta_+^{} + (12.10-0.23 i) \Delta_+^{} \big] \rho_+^{}
\vspace{1pt} \\ +\; ( 0.02\, \delta_-^{} + 0.01\, \Delta_-^{} ) \rho_-^{} \,, \end{array}
\nonumber \\ \vphantom{|^{\int_|^\int}}
{\cal A}_{\bar B_s\to\eta'\omega}^{(1)} & \,\simeq\, 0.05 - 0.33 i
- \big[ (1.22 - 0.02 i) \delta_+^{} + (2.44 - 0.05 i) \Delta_+^{} \big] \rho_+^{} \,,
\nonumber \\ \vphantom{|^{\int_\int^\int}}
{\cal A}_{\bar B_s\to\eta\omega}^{(2)} & \,\simeq\, 0.52 + 0.74 i
+ \big[ (5.83 - 0.11 i) \delta_+^{} + (11.70 - 0.22 i) \Delta_+^{} \big] \rho_+^{}
+ 0.01\, \Delta_-^{} \rho_-^{} \,,
\nonumber \\ \vphantom{|^{\int_|^\int}}
{\cal A}_{\bar B_s\to\eta'\omega}^{(2)} & \,\simeq \begin{array}[t]{l} 3.23 - 3.83 i
- \big[ (1.37 - 0.03 i) \delta_+^{} + (2.72 - 0.05 i) \Delta_+^{} \big] \rho_+^{}
\vspace{1pt} \\ -\; ( 0.05\, \delta_-^{} + 0.01\, \Delta_-^{} ) \rho_-^{} \,, \end{array}
\nonumber \\ \vphantom{|^{\int_\int^\int}}
{\cal A}_{\bar B_s\to\phi\omega} & \,\simeq \begin{array}[t]{l} -1.69 - 1.41 i
- \big[ (14.00 - 0.26 i) \delta_+^{} + (28.10 - 0.53 i) \Delta_+^{} \big] \rho_-^{}
\vspace{1pt} \\ -\; ( 0.01\, \delta_-^{} + 0.03\, \Delta_-^{} ) \rho_+^{} \,, \end{array}
\end{align}
where the superscripts (1) and (2) refer to SCET Solutions 1 and 2, respectively,
$Z'$ terms with numerical factors below 0.005 in size are not displayed, and
\begin{align} \label{dDr}
\delta_{\pm}^{} & \,=\, \delta_L^{} \pm \delta_R^{} \,, &
\Delta_{\pm}^{} & \,=\, \Delta_L^{dd} \pm \Delta_R^{dd} \,, &
\rho_{\pm}^{} & \,=\, \rho_L^{} \pm \rho_R^{} \,. &
\end{align}

Given that $\delta_{\pm}^{}$ and $\rho_{\pm}^{}$ participate in the amplitudes for
\,$\bar B_s^0\to(\eta,\eta',\phi)(\pi^0,\rho^0)$\, as well \cite{Faisel:2017glo}, as 
Eqs.\,\,(\ref{eta'rho1})-(\ref{phirho}) in the Appendix show, 
it is germane to include them in this analysis.
What is more, as discussed in Ref.\,\,\cite{Faisel:2017glo}, the LHCb finding
\,${\cal B}\big(\bar B_s\to\phi\rho^0\big){}_{\rm exp}^{}=(0.27\pm0.08)\times10^{-6}$\,
\cite{pdg,Aaij:2016qnm}, which is in line with some of its SM estimates albeit within large
errors\,\,\cite{Ali:2007ff,Cheng:2009mu,Yan:2018fif,Wang:2017rmh,Faisel:2011kq,Hofer:2010ee},
translates into an important constraint on the $Z'$ couplings.
Additionally, treating all these rare decays at the same time would allow us to see if there 
might be correlations among their rate increases/decreases compared to the SM expectations.
Such correlations would constitute $Z'$ predictions potentially testable in
upcoming experiments.

The couplings in Eq.\,(\ref{dDr}) also affect other nonleptonic \,$b\to s$\, processes which 
have been observed and hence need to respect the restrictions implied by their data.
Here we focus on the well-measured decays \,$B^-\to\pi^0K^-,\pi^-\bar K^0$\, and 
\,$\bar B^0\to\pi^0\bar K^0,\pi^+K^-$\, plus their antiparticle counterparts.
Their rates in the SM, with {\footnotesize{$\sim$\,}}40\% errors \cite{Williamson:2006hb}, 
agree with their measurements~\cite{pdg}.
Incorporating the $Z'$ terms, we have calculated the $B\to\pi K$ amplitudes and collected their 
expressions in Eqs.\,\,(\ref{piKbar}) and\,\,(\ref{piK}).

To illustrate how the $Z'$ interactions contribute to the decays of interest, we put together 
5,000 randomly generated benchmarks fulfilling the following conditions.
We impose
\begin{equation} \label{constr1}
0.11 \,\le\, 10^6\,{\cal B}\big(\bar B_s\to\phi\rho^0\big) \,\le\, 0.43 \,,
\end{equation}
which is the 2$\sigma$ range of ${\cal B}\big(\bar B_s\to\phi\rho^0\big){}_{\rm exp}^{}$.
For the $B\to\pi K$ requirement, since the SM predictions have uncertainties of around 40\% and are 
compatible with their data, we demand that the $Z'$ effects alter the \,$B\to\pi  K$\, rates relative 
to their SM values by no more than 20\%.\footnote{The $CP$ asymmetry difference 
\,$\Delta A_{CP}^{\rm exp} = 
A_{CP}^{\rm exp}(B^-\to\pi^0 K^-)-A_{CP}^{\rm exp}(\bar B^0\to\pi^+K^-)=0.122\pm0.022$ \cite{pdg} 
might be another restraint.
It excludes the SM central values \,$\Delta A_{CP}^{\rm SM}\simeq-0.01$\, for the two SCET solutions, 
but the theoretical error is large, ${\scriptstyle\sim\,}0.15$ \cite{Williamson:2006hb},
implying that the predictions are still consistent with $\Delta A_{CP}^{\rm exp}$.
Similarly, although our $Z'$ benchmarks yield 
\,$-0.025\,{\scriptstyle\lesssim}\,\Delta A_{CP}^{{\rm SM}+Z'\!}{\scriptstyle\lesssim}\,0.002$,\, 
they are not in conflict with $\Delta A_{CP}^{\rm exp}$, considering the substantial theoretical 
uncertainty.}
For the $Z'$ parameters, we select \,$\rho_R^{}=0.1\,\rho_L^{}$\, as in Ref.\,\cite{Faisel:2017glo} 
and let the products of $\rho_L^{}$ and the other nonleptonic couplings vary within the intervals
\begin{equation}
\delta_{\pm\,}^{} \rho_L^{} \,\in\, [-1,1] \,, ~~~ ~~~~
\Delta_{\pm\,}^{} \rho_L^{} \,\in\, [-1,1] \,,
\end{equation}
having already set \,$m_{Z'}^{}=1$\,TeV\, in Eq.\,(\ref{ampso}) and in the other amplitudes 
written down in the Appendix.
We present the results in the two figures below which depict two-dimensional projections of 
the benchmarks for a number of quantities.

\begin{figure}[t]
\includegraphics[height=55mm]{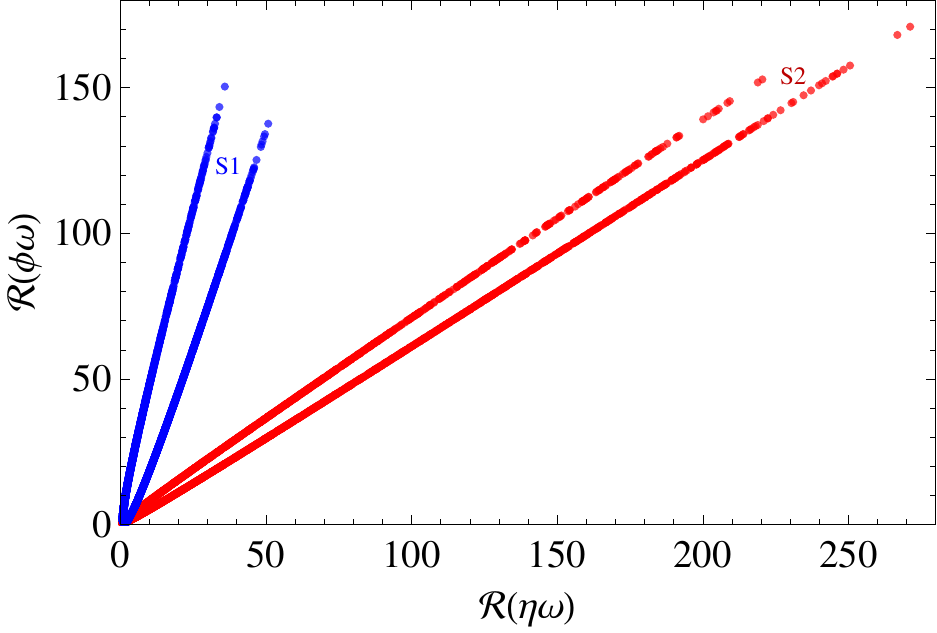} ~ ~ ~~
\includegraphics[height=55mm]{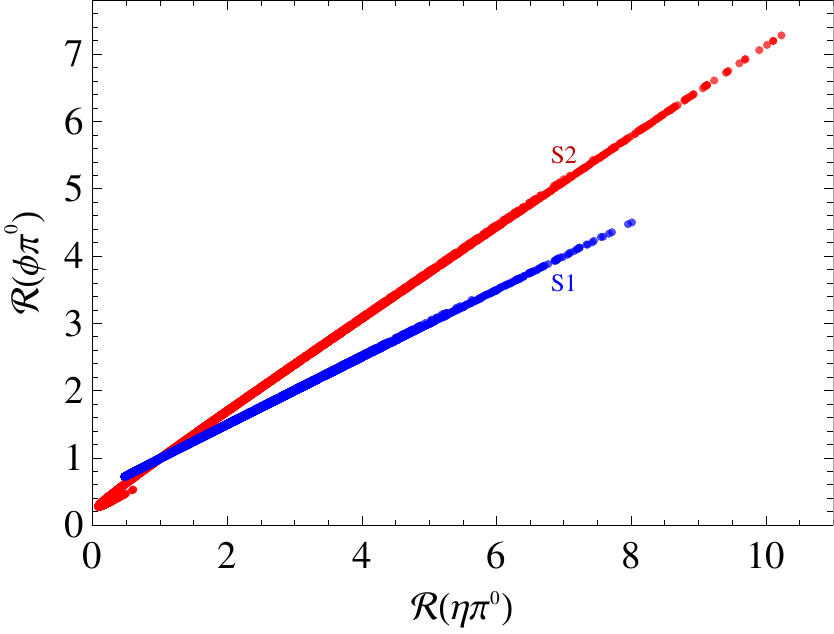}\vspace{1ex}\\ ~~~~
\includegraphics[width=77mm]{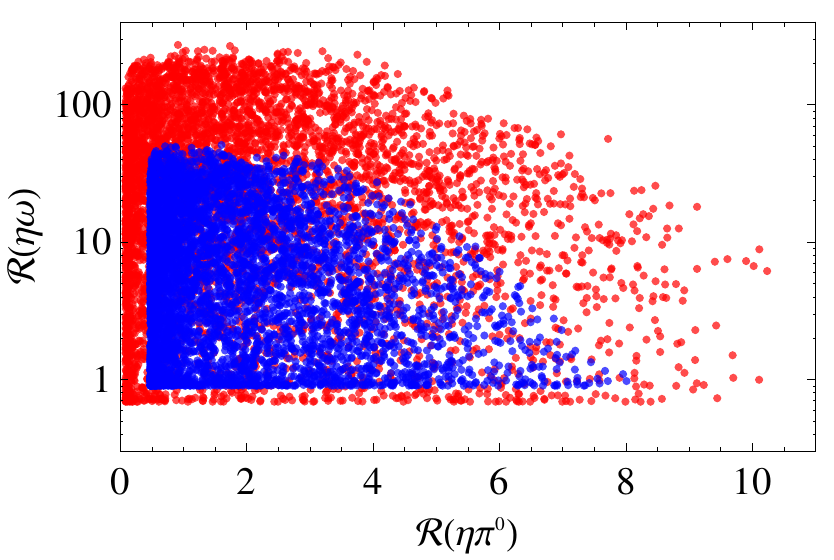} ~~~
\includegraphics[width=77mm]{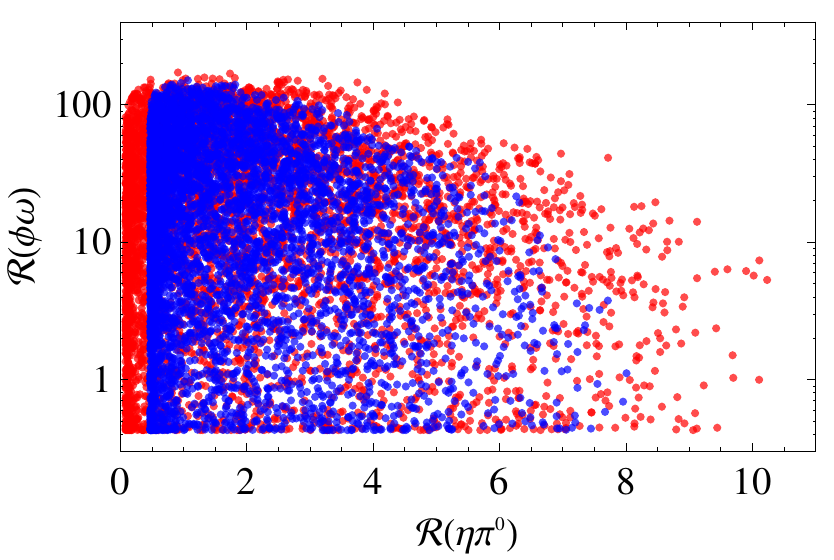}\vspace{-7pt}
\caption{Distributions of the ratio
\,${\cal R}(MM')=\Gamma_{\bar B_s\to MM'}/\Gamma_{\bar B_s\to MM'}^{\textsc{sm}}$\, among different 
pairs of final states $MM'$ for the benchmarks corresponding to SCET Solutions 1 (blue, S1) 
and 2 (red, S2)} \label{Rs}
\end{figure}

In Fig.\,\ref{Rs} we display the distributions of the enhancement factor
\begin{equation}
{\cal R}(MM') \,=\, \frac{\Gamma_{\bar B_s\to MM'}^{}}{\Gamma_{\bar B_s\to MM'}^{\textsc{sm}}}
\end{equation}
of the \,$\bar B_s\to MM'$\, rate with respect to its SM prediction for a few pairs of final states $MM'$.
As the top-left plot reveals, ${\cal R}(\eta\omega)$ and ${\cal R}(\phi\omega)$ go up or down
simultaneously and can reach roughly 50 and 150 (270 and 170), respectively, for Solution 1 (2).
It follows that, in light of the SCET central values in Table\,\,\ref{smbf}, the $Z'$ influence
can boost the branching fractions of \,$\bar B_s\to\eta\omega$\, and \,$\bar B_s\to\phi\omega$\,
to \,{\footnotesize$\sim$\,}$2\times10^{-6}$ and {\footnotesize$\sim$\,}$6\times10^{-6}$,\,
respectively, for both solutions.
Accordingly, these decay channels are potentially sensitive to NP signals, and moreover the correlation 
between ${\cal R}(\eta\omega)$ and ${\cal R}(\phi\omega)$ may be experimentally checked.

As regards \,$\bar B_s\to\eta'\omega$,\, for which we do not provide any graphs, with Solution 1 (2) 
we get at most \,${\cal R}(\eta'\omega)\sim80$ (only 2.5), which translates into
\,${\cal B}(\bar B_s\to\eta'\omega)\,\mbox{\footnotesize$\lesssim$}\,0.08\,(0.5)\times10^{-6}$\, 
based on its SCET central values in Table\,\,\ref{smbf}.
However, the corresponding upper error in this table for Solution 1 suggests that 
${\cal B}(\bar B_s\to\eta'\omega)$ might still undergo a $Z'$-mediated boost to the $10^{-6}$ level, 
thereby offering an additional window to the $Z'$ interactions.

The top-right plot in Fig.\,\ref{Rs} indicates that ${\cal R}(\eta\pi^0)$ and ${\cal R}(\phi\pi^0)$, 
like ${\cal R}(\eta\omega)$ and ${\cal R}(\phi\omega)$, increase/decrease at the same time, although 
the former two can rise to only about 8.0 and 4.5 (10 and 7.3), respectively, for Solution 1 (2).
Nevertheless, as elaborated in Ref.\,\,\cite{Faisel:2017glo}, such enhancement factors are 
sufficiently sizable to make \,$\bar B_s\to(\eta,\phi)\pi^0$\, promising as extra tools in the quest 
for the potential NP behind the \,$b\to s\mu^+\mu^-$\, anomalies.
The correlation between ${\cal R}(\eta\pi^0)$ and ${\cal R}(\phi\pi^0)$ is obviously a testable 
prediction 
as well.
We notice that the preceding Solution-2 numbers are roughly 20\% less than their counterparts 
(12 and 9.1) in Ref.\,\,\cite{Faisel:2017glo}, mostly due to the aforementioned $B\to\pi K$ requisite.

From the bottom plots in Fig.\,\ref{Rs}, unlike the top ones, it is not evident if there is 
a connection between ${\cal R}(\eta\omega)$ or ${\cal R}(\phi\omega)$ and ${\cal R}(\eta\pi^0)$.
The former two also do not seem to have clear correlations with ${\cal R}(\phi\pi^0)$, although this 
is not illustrated here.
We will ignore possibly related consequences for \,$\bar B_s^0\to\eta'\pi^0,(\eta,\eta')\rho^0$\, 
because the $Z'$ impact on their rates is only modest \cite{Faisel:2017glo}.

\begin{figure}[b]
\includegraphics[width=41mm]{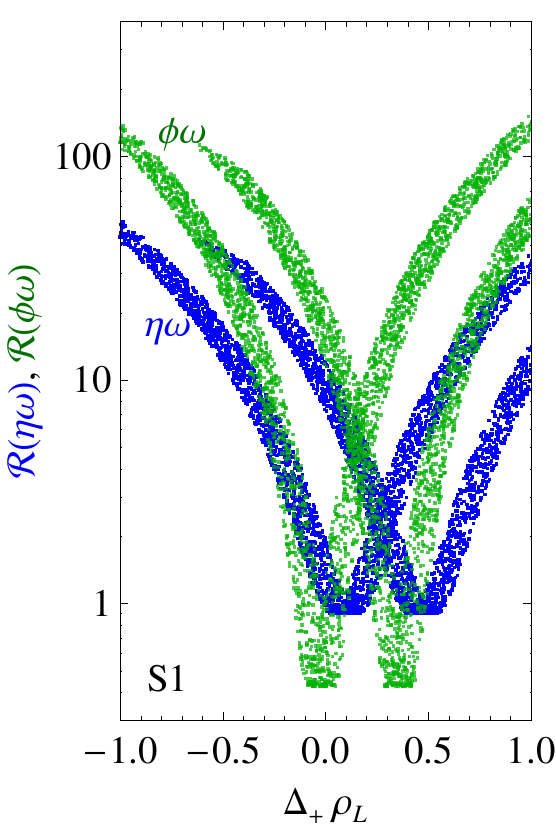} \!\!\!
\includegraphics[width=41mm]{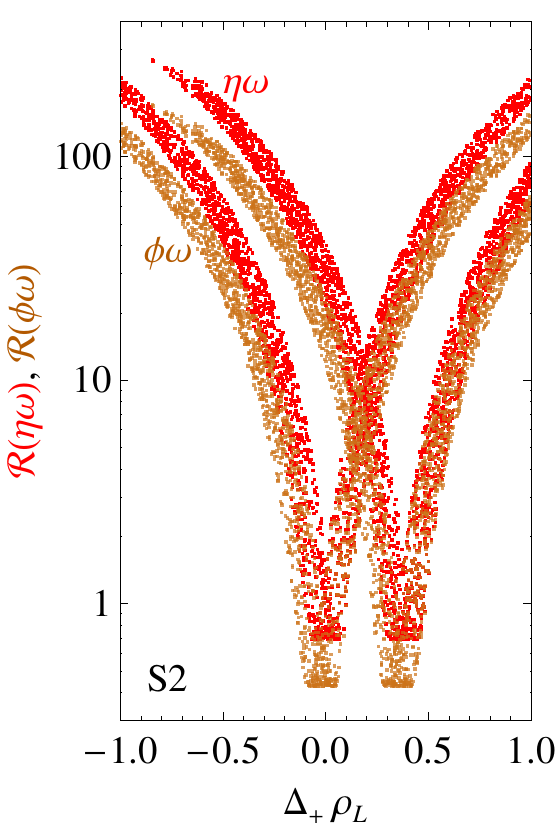} ~ ~
\includegraphics[width=41mm]{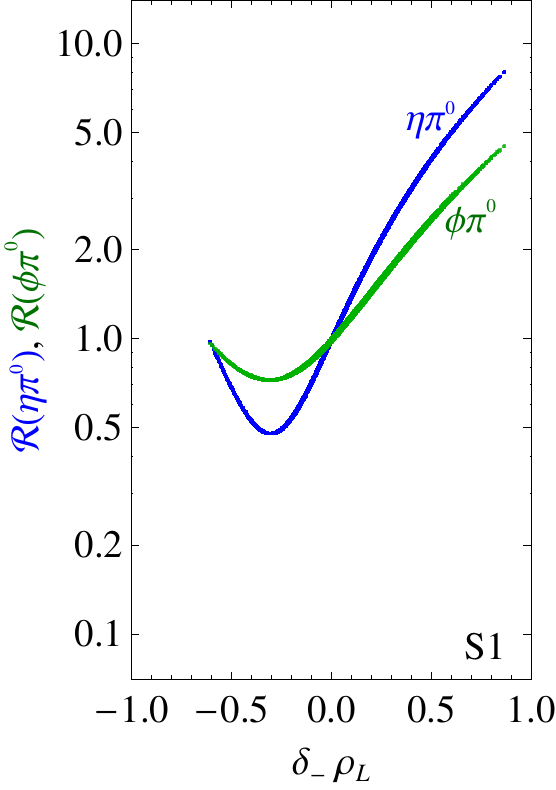} \!\!\!
\includegraphics[width=41mm]{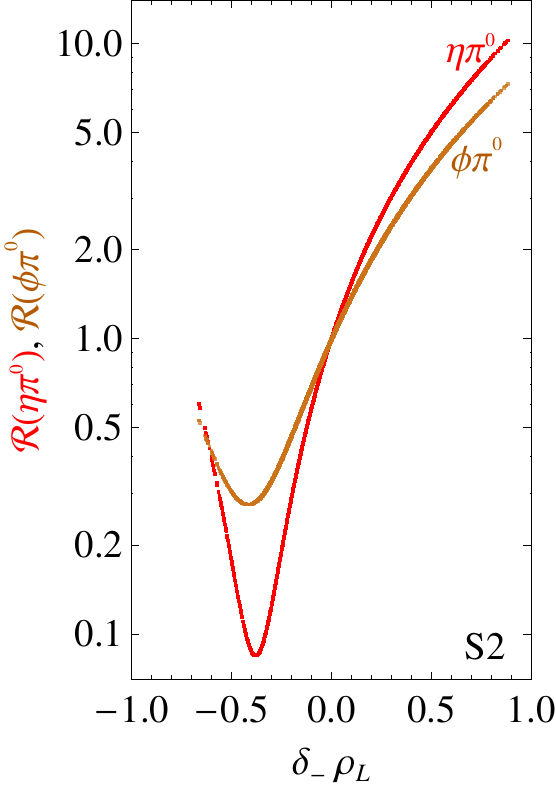}\vspace{-7pt}
\caption{Distributions of ${\cal R}(\eta\omega)$ and ${\cal R}(\phi\omega)$ versus 
$\Delta_+^{}\rho_L^{}$ and of ${\cal R}(\eta\pi^0)$ and ${\cal R}(\phi\pi^0)$ versus 
$\delta_-^{}\rho_L^{}$ for SCET Solutions 1 (S1) and 2 (S2).} \label{R-coupling}
\end{figure}

Information about relationships between ${\cal R}(MM')$ and the $Z'$ couplings is highly valuable 
for examining the latter if one or more of these decays are observed.
For our decay channels of greatest interest, it turns out that there are a few relationships that 
are more or less plain, which we exhibit in Fig.\,\ref{R-coupling}.
As might be expected, the curves in the fourth plot resemble the corresponding ones in 
Ref.\,\cite{Faisel:2017glo}.

\begin{figure}[b!] \medskip
\includegraphics[height=59mm]{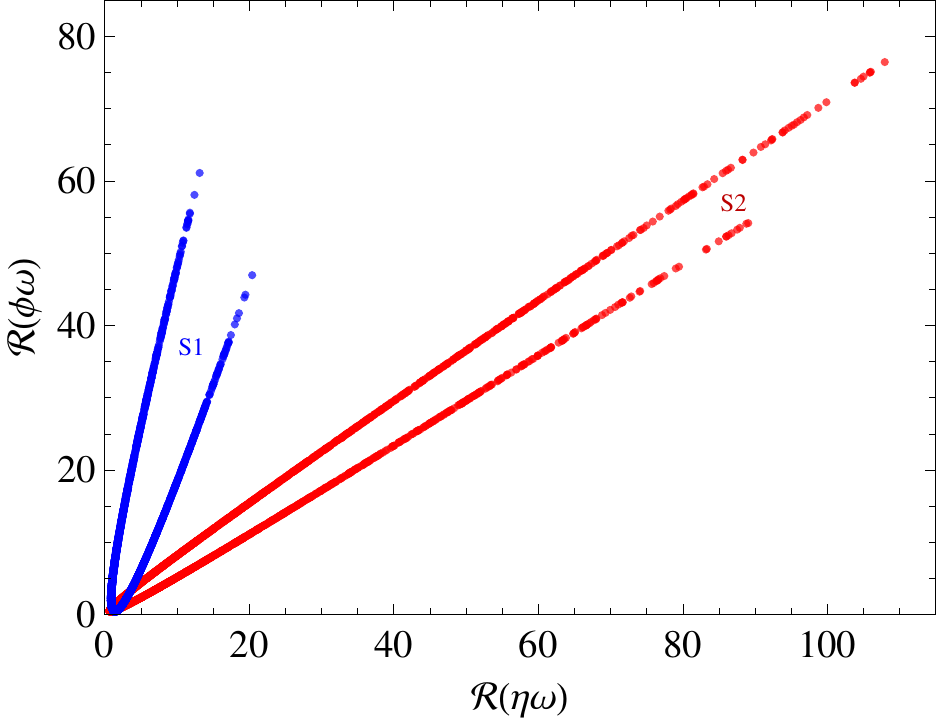} ~ ~ ~
\includegraphics[height=59mm]{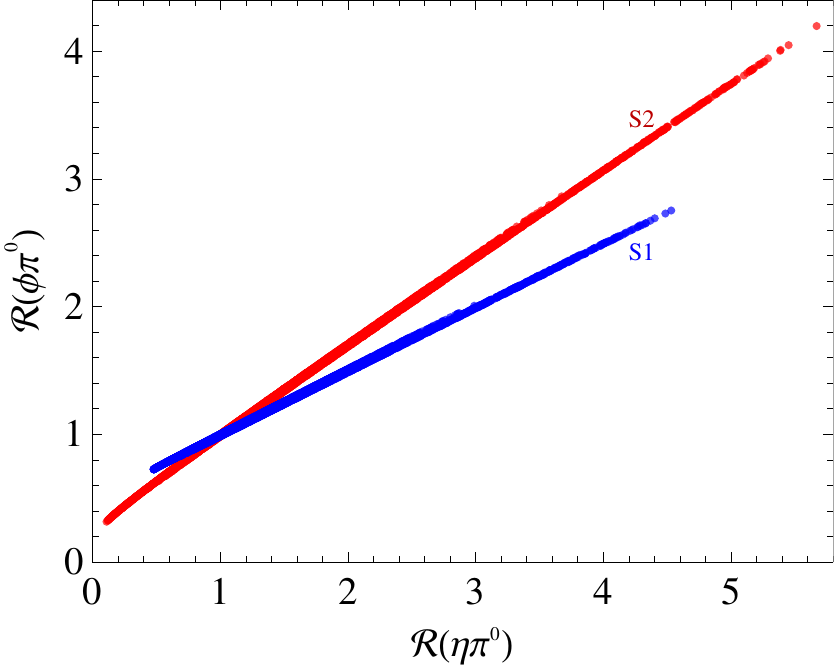}\vspace{-7pt}
\caption{The same as the top two plots in Fig.\,\ref{Rs}, but with the stronger requirement that 
the $Z'$-induced modifications to the $B\to\pi  K$ rates in the SM not exceed 10\%.} \label{Rs10pc}
\end{figure}

\begin{figure}[t!] \bigskip
\includegraphics[height=53mm]{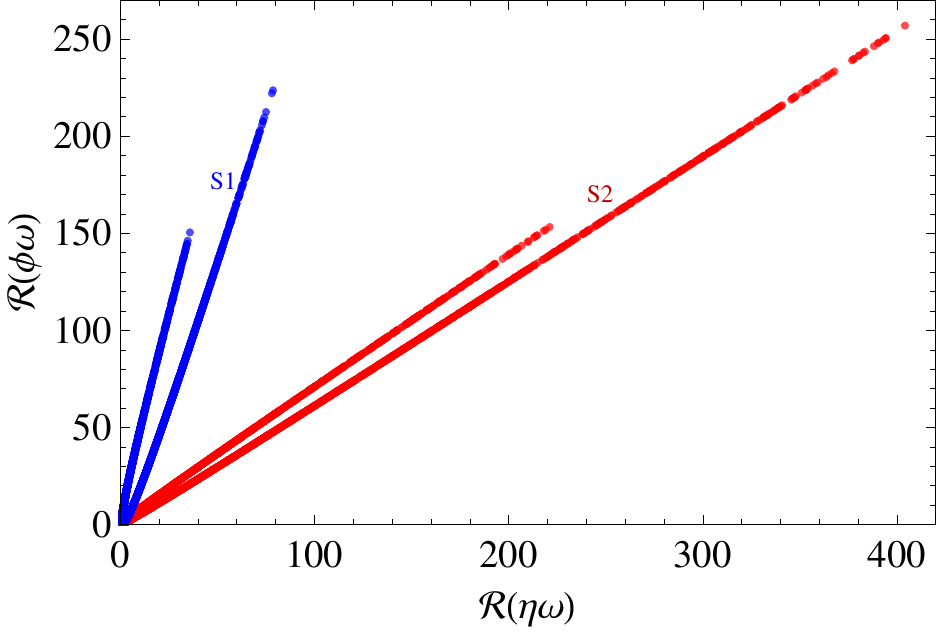} ~ ~ ~
\includegraphics[height=53mm]{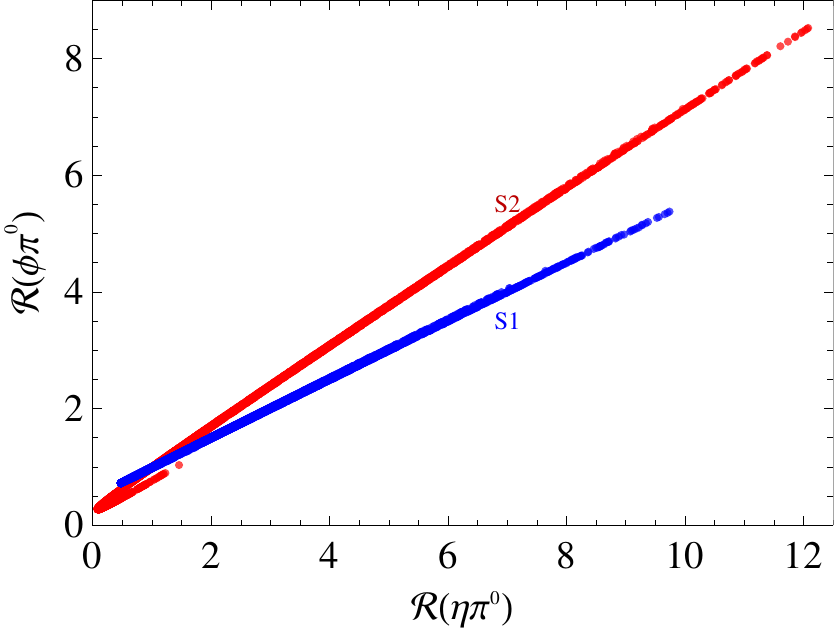}\vspace{-7pt}
\caption{The same as Fig.\,\ref{Rs10pc}, but with the weaker requirement that the $Z'$-induced 
modifications to the \,$B\to\pi K$\, rates in the SM not exceed 25\%.} \label{Rs30pc}
\end{figure}

It is worth noting that the restriction we imposed above from the $B\to\pi K$ sector is of 
significance to some extent, although how stringent the condition should be is unclear in view of 
the 40\% uncertainties of the SM rate predictions \cite{Williamson:2006hb}.
For illustration, making it stricter so that the $Z'$-induced changes to the $B\to\pi  K$ rates in 
the SM not exceed 10\%, we arrive at the graphs in Fig.\,\ref{Rs10pc}.
In this instance, ${\cal R}(\eta\pi^0)$ and ${\cal R}(\phi\pi^0)$, especially the latter, 
become somewhat less remarkable than before, but ${\cal R}(\eta\omega)$ and ${\cal R}(\phi\omega)$ 
are still considerable, and so all these decays remain useful for probing the $Z'$ effects.
We further find, however, that if the $B\to\pi K$ rates are allowed to deviate from the SM 
expectations by up to 25\% or higher, the impact on the maxima of these ${\cal R}$s will start to 
diminish and they can have the wider spreads depicted in Fig.\,\ref{Rs30pc}.

Lastly, we comment that the $Z'$ couplings in our benchmarks are consistent with collider 
constraints, as discussed in Ref.\,\,\cite{Faisel:2017glo}, including those from LHC searches 
for new high-mass phenomena in the dilepton final states \cite{Aaboud:2017buh}.
This is partly because the products of the muon-$Z'$ coupling $\Delta_V^{\mu\mu}$ and the quark-$Z'$ 
flavor-diagonal couplings $(\delta_\pm,\Delta_\pm)$ can be rendered small enough to evade the bounds by sufficiently increasing the size of $\rho_L^{}$ while maintaining $\Delta_V^{\mu\mu}\rho_L^{}$ in Eq.\,(\ref{c9c9'}) and $(\delta_\pm,\Delta_\pm)\rho_L^{}$ to stay within their desired respective ranges.

\section{Conclusions\label{conclusions}}

We have explored the possibility that the anomalies detected in the current \,$b\to s\mu^+\mu^-$\, data arise from physics beyond the SM and that the same new physics also affects the rare nonleptonic decays of the $\bar B_s$ meson, most of which are not yet observed.
Since the rates of these modes in the SM are comparatively low, one or more of them may be sensitive to NP signals.
Adopting a scenario in which the NP is due to the interactions of a heavy $Z'$ boson, we investigate the implications for the rare decays \,$\bar B_s^0\to(\eta,\eta',\phi)(\pi^0,\rho^0,\omega)$.\,
Taking into account the pertinent restraints, we demonstrate that the $Z'$ effects can hugely amplify the rates of \,$\bar B_s^0\to(\eta,\phi)\omega$\, above their SM predictions, by as much as two orders of magnitude.
The corresponding enhancement factors for \,$\bar B_s^0\to(\eta,\phi)\pi^0$\, could be substantial as well, up to an order of magnitude, in line with our previous work.
Thus, these four rare modes are potentially very consequential should future experiments establish that the \,$b\to s\mu^+\mu^-$\, anomalies are really manifestations of NP.

\acknowledgements

The work of J.T was supported in part by the Republic of China Ministry of Education
Academic Excellence Program (Grant No. 105R891505).

\appendix

\section{Additional amplitudes\label{amps}}

We have derived the main formulas for the $Z'$ contributions to
\,$\bar B_s\to(\eta,\eta',\phi)(\pi^0,\rho^0)$\, in Ref.\,\cite{Faisel:2017glo} under the SCET framework.
Therein we did not include Solution 1 in the evaluation of the $Z'$ effects and neglected renormalization-group evolution for simplicity.
In the present paper, we include the latter and give results for both Solutions 1 and 2.
Thus, summing the SM and $Z'$ terms, with the central values of the input parameters and
\,$m_{Z'}^{}=1$\,TeV,\, we calculate the amplitudes to be, in units of $10^{-9}$ GeV,
\begin{align}
{\cal A}_{\bar{B}_s\to\eta\pi^0}^{(1)} &\,\simeq\, 1.43 + 1.31 i
+ 0.03\, \delta_+^{} \rho_-^{} + (4.15-0.08 i) \delta_-^{} \rho_+^{} \,,
\nonumber \\
{\cal A}_{\bar{B}_s\to\eta'\pi^0}^{(1)} &\,\simeq\, -0.31 - 0.21 i
- (0.84 - 0.02 i) \delta_-^{} \rho_+^{} \,,
\nonumber \\
{\cal A}_{\bar{B}_s\to\phi\pi^0}^{(1)} &\,\simeq\, -1.63 - 2.53 i
- (5.57 - 0.11 i) \delta_-^{} \rho_-^{} - 0.08\, \delta_+^{} \rho_+^{} \,,
\nonumber \\
{\cal A}_{\bar{B}_s\to\eta\rho^0}^{(1)} &\,\simeq\, 2.18 + 2.17 i
+ 0.06\, \delta_-^{} \rho_-^{} + (6.60-0.13 i)\delta_+^{} \rho_+^{} \,,
\nonumber \\ \label{eta'rho1}
{\cal A}_{\bar{B}_s\to\eta'\rho^0}^{(1)} &\,\simeq\, -0.47 - 0.34 i
- 0.01\, \delta_-^{} \rho_-^{} - (1.33 - 0.03 i)\delta_+^{} \rho_+^{} \,,
\end{align}
\begin{align}
{\cal A}_{\bar{B}_s\to\eta\pi^0}^{(2)} & \,\simeq\, 1.67 + 0.47 i
+ (3.98 - 0.08 i) \delta_-^{} \rho_+^{} \,,
\nonumber \\
{\cal A}_{\bar{B}_s\to\eta'\pi^0}^{(2)} & \,\simeq\, 0.48 - 2.48 i
- 0.09\, \delta_+^{} \rho_-^{} - (1.00 - 0.02 i) \delta_-^{} \rho_+^{} \,,
\nonumber \\
{\cal A}_{\bar{B}_s\to\phi\pi^0}^{(2)} & \,\simeq\, -2.88 - 1.69 i
- (7.58 - 0.15 i) \delta_-^{} \rho_-^{} - 0.03\, \delta_+^{} \rho_+^{} \,,
\nonumber \\
{\cal A}_{\bar{B}_s\to\eta\rho^0}^{(2)} & \,\simeq\, 2.56 + 0.77 i
+ (6.35 - 0.12 i) \delta_+^{} \rho_+^{} \,,
\nonumber \\ \label{eta'rho2}
{\cal A}_{\bar{B}_s\to\eta'\rho^0}^{(2)} & \,\simeq\, 0.78 - 4.12 i
- 0.16\, \delta_-^{} \rho_-^{} - (1.52 - 0.03 i) \delta_+^{} \rho_+^{} \,,
\\ \vphantom{|^{\int_\int^\int}} \label{phirho}
{\cal A}_{\bar{B}_s\to\phi\rho^0} & \,\simeq\, -6.53 - 1.47 i
- (15.50 - 0.29 i) \delta_+^{} \rho_-^{} + 0.01\, \delta_-^{} \rho_+^{} \,,
\end{align}
where the superscripts (1) and (2) refer to Solutions 1 and 2, respectively, $Z'$ terms with numerical factors below 0.005 are not displayed, and $\delta_\pm^{}$ and $\rho_\pm^{}$ were already
defined in Eq.\,(\ref{dDr}). We note that in SCET at leading order the rates of \,$\bar B_s^0\to(\eta,\eta',\phi)(\pi^0,\rho^0,\omega)$\, are equal to their antiparticle counterparts
\cite{Wang:2017rmh,Wang:2008rk,Williamson:2006hb}.
We have checked that under the same requirements as in Ref.\,\cite{Faisel:2017glo} the resulting maximal enhancement factor of the \,$\bar B_s^0\to\eta\pi^0\,\big(\phi\pi^0\big)$ rate is around 12 (8.6), which is almost identical to (5\% below) what we determined earlier \cite{Faisel:2017glo} ignoring RGE.
Imposing also the $B\to\pi  K$ condition as discussed in Sec.\,\ref{Bs2MM'} may lead to smaller enhancement factors, depending on how stringent it is.

In the SCET approach the SM amplitudes for the $B\to\pi  K$ channels are dominated by the so-called charming-penguin terms.
The relevant hard kernels are available from Ref.\,\,\cite{Williamson:2006hb}, with the quantities $c_{1,2,3,4}$ and $b_{1,2,3,4}$ now involving \,$C_j^-=C_j^{}-\tilde C_j^{}$,\, analogously to the \,$\bar B_s\to PP'$\, case.
Including the $Z'$ parts similarly to the previous paragraph, with Solution 1 we obtain, in units of $10^{-9}$ GeV,
\begin{align}
{\cal A}_{B^-\to\pi^0K^-}^{(1)} \,\simeq\, &~ \mbox{$-35.10 + 1.45 i$}
- \big[ (3.32 - 0.06 i) \delta_+^{} + (3.19-0.06i) \Delta_+^{} \big] \rho_-^{}
\nonumber \\ & - \big[ (5.24 - 0.10 i) \delta_-^{} + 0.21\, \Delta_-^{} \big] \rho_+^{} \,,
\nonumber \\ \vphantom{|^{\int_|^\int}}
{\cal A}_{B^-\to\pi^-\bar K^0}^{(1)} \,\simeq\, &~ \mbox{$-47.10 + 9.63 i$}
+ \big[ 0.06\, \delta_+^{}-(4.51-0.09 i) \Delta_+^{} \big] \rho_-^{}
\nonumber \\ & + \big[ 0.02\, \delta_-^{} - (0.30-0.01 i) \Delta_-^{} \big] \rho_+^{} \,,
\nonumber \\ \vphantom{|^{\int_|^\int}}
{\cal A}_{\bar B^0\to\pi^0\bar K^0}^{(1)} \,\simeq\, &~ 31.60 - 8.40 i
- \big[ 0.08\, \delta_+^{} - (3.19-0.06 i) \Delta_+^{} \big] \rho_-^{}
\nonumber \\ & - \big[ (5.01-0.10 i) \delta_-^{} - 0.21\, \Delta_-^{} \big] \rho_+^{} \,,
\nonumber \\ \vphantom{|^{\int_|^\int}}
{\cal A}_{\bar B^0\to\pi^+K^-}^{(1)} \,\simeq\, &~ \mbox{$-47.30 + 4.30 i$}
- \big[ (4.63-0.09 i) \delta_+^{} + (4.51-0.09 i) \Delta_+^{} \big] \rho_-^{}
\nonumber \\ & - \big[ (0.33-0.01 i)\delta_-^{} + (0.30-0.01 i)\Delta_-^{} \big] \rho_+^{} \,,
\end{align}
\begin{align}
{\cal A}_{B^+\to\pi^0K^+}^{(1)} \,\simeq\, &~ \mbox{$-35.10 + 11.90 i$}
- \big[ (2.94+0.06 i) \delta_+^{} + (2.81+0.05 i) \Delta_+^{} \big] \rho_-^{}
\nonumber \\ & - \big[ (4.86+0.09 i) \delta_-^{} - 0.17\, \Delta_-^{} \big] \rho_+^{} \,,
\nonumber \\ \vphantom{|^{\int_|^\int}}
{\cal A}_{B^+\to\pi^+K^0}^{(1)} \,\simeq\, &~ \mbox{$-47.60 + 9.15 i$}
+ \big[ 0.06\, \delta_+^{}-(3.97+0.08 i) \Delta_+^{} \big] \rho_-^{}
+ \big( 0.02\, \delta_-^{} + 0.24\, \Delta_-^{} \big) \rho_+^{} \,,
\nonumber \\ \vphantom{|^{\int_|^\int}}
{\cal A}_{B^0\to\pi^0K^0}^{(1)} \,\simeq\, &~ 32.00 - 4.89 i
- \big[ 0.09\, \delta_+^{} - (2.81+0.05 i) \Delta_+^{} \big] \rho_-^{}
\nonumber \\ & - \big[ (5.02-0.10 i) \delta_-^{} + 0.17\, \Delta_-^{} \big] \rho_+^{} \,,
\nonumber \\ \vphantom{|^{\int_|^\int}}
{\cal A}_{B^0\to\pi^-K^+}^{(1)} \,\simeq\, &~ \mbox{$-47.20 + 14.60 i$}
- \big[ (4.10+0.08 i) \delta_+^{} + (3.97+0.08 i) \Delta_+^{} \big] \rho_-^{}
\nonumber \\ & + \big( 0.20\, \delta_-^{} + 0.24\, \Delta_-^{} \big) \rho_+^{}
\end{align}
and with Solution 2
\begin{align}
{\cal A}_{B^-\to\pi^0K^-}^{(2)} \,\simeq\, &~ \mbox{$-35.10 + 2.04 i$}
- \big[ (2.96-0.06 i) \delta_+^{} + (2.85-0.05 i) \Delta_+^{} \big] \rho_-^{}
- (4.88 - 0.09 i) \delta_-^{} \rho_+^{} \,,
\nonumber \\ \vphantom{|^{\int_|^\int}}
{\cal A}_{B^-\to\pi^-\bar K^0}^{(2)} \,\simeq\, &~ \mbox{$-47.40 + 9.95 i$}
+ \big[ 0.06\, \delta_+^{} - (4.02-0.08 i) \Delta_+^{} \big] \rho_-^{}
+ 0.02\, \delta_-^{} \rho_+^{} \,,
\nonumber \\ \vphantom{|^{\int_|^\int}}
{\cal A}_{\bar B^0\to\pi^0\bar K^0}^{(2)} \,\simeq\, &~ 31.70 - 8.38 i
- \big[ 0.07\, \delta_+^{} - (2.85-0.05 i) \Delta_+^{} \big] \rho_-^{}
- (4.87-0.09 i) \delta_-^{} \rho_+^{} \,,
\nonumber \\ \vphantom{|^{\int_|^\int}} \label{piKbar}
{\cal A}_{\bar B^0\to\pi^+K^-}^{(2)} \,\simeq\, &~ \mbox{$-47.20 + 4.77 i$}
- \big[ (4.14-0.08 i) \delta_+^{} + (4.02-0.08 i) \Delta_+^{} \big] \rho_-^{}
- 0.03\, \delta_-^{} \rho_+^{} \,,
\end{align}
\begin{align}
{\cal A}_{B^+\to\pi^0K^+}^{(2)} \,\simeq\, &~ \mbox{$-35.10 + 11.80 i$}
- \big[ (2.65+0.05 i) \delta_+^{} + (2.54+0.05 i) \Delta_+^{} \big] \rho_-^{}
\nonumber \\ & - \big[ (4.57+0.09 i) \delta_-^{} - (0.31+0.01 i) \Delta_-^{} \big] \rho_+^{} \,,
\nonumber \\ \vphantom{|^{\int_|^\int}}
{\cal A}_{B^+\to\pi^+K^0}^{(2)} \,\simeq\, &~ \mbox{$-47.80 + 9.49 i$}
+ \big[ 0.06\, \delta_+^{}-(3.59+0.07 i) \Delta_+^{} \big] \rho_-^{}
\nonumber \\ & + \big[ 0.02\, \delta_-^{} + (0.44+0.01 i) \Delta_-^{} \big] \rho_+^{} \,,
\nonumber \\ \vphantom{|^{\int_|^\int}}
{\cal A}_{B^0\to\pi^0K^0}^{(2)} \,\simeq\, &~ 32.00 - 5.37 i
- \big[ 0.07\, \delta_+^{} - (2.54+0.05 i) \Delta_+^{} \big] \rho_-^{}
\nonumber \\ & - \big[ (4.87+0.09 i) \delta_-^{} + (0.31+0.01 i) \Delta_-^{} \big] \rho_+^{} \,,
\nonumber \\ \vphantom{|^{\int_|^\int}} \label{piK}
{\cal A}_{B^0\to\pi^-K^+}^{(2)} \,\simeq\, &~ \mbox{$-47.20 + 14.80 i$}
- \big[ (3.71+0.07 i) \delta_+^{} + (3.59+0.07 i) \Delta_+^{} \big] \rho_-^{} \hspace{7em}
\nonumber \\ & + \big[ (0.40+0.01 i) \delta_-^{} + (0.44+0.01 i) \Delta_-^{} \big] \rho_+^{} \,.
\end{align}
In the SM limit, for Solution 1\,(2) these amplitudes yield the $CP$-averaged
branching fractions  ${\cal B}\big(B^+\to\pi^0K^+)_{\textsc{sm}}=12.1\, (12.1)$,
\,${\cal B}\big(B^+\to\pi^+K^0)_{\textsc{sm}}=21.7\, (21.9)$,
\,${\cal B}\big(B^0\to\pi^0K^0)_{\textsc{sm}}= 9.1\, ( 9.2)$, and
\,${\cal B}\big(B^0\to\pi^-K^+)_{\textsc{sm}}=20.3\, (20.3)$,\,
all in units of $10^{-6}$, with uncertainties of about 40\% \cite{Williamson:2006hb}.
In view of the errors, these predictions are compatible with their experimental
counterparts \cite{pdg}  ${\cal B}\big(B^+\to\pi^0K^+)=12.9\pm0.5$,
\,${\cal B}\big(B^+\to\pi^+K^0)=23.7\pm0.8$,
\,${\cal B}\big(B^0\to\pi^0K^0)= 9.9\pm0.5$,\, and
\,${\cal B}\big(B^0\to\pi^-K^+)=19.6\pm0.5$,\, all in units of $10^{-6}$.

\end{document}